\documentclass[11pt,twoside]{article}

\pdfoutput=1
\usepackage{asp2006}
\usepackage{graphicx}

\begin{document}
\title{Diffuse microwave emission of the interstellar medium : intensity and polarization}   
\author{Marc-Antoine Miville-Desch\^enes}   
\affil{Institut d'Astrophysique Spatiale (CNRS), Universit\'e Paris-Sud, b\^atiment 121, 91405, Orsay, France }

\begin{abstract} 
With the increasing number of experiments dedicated to the observation of the Cosmic Microwave Background,
interest on the detailed properties of the foreground Galactic emission has risen. 
In particular the focus is shifting towards polarized microwave emission 
with the goal of detecting signature from primordial gravitational waves and inflation. 
This review describes the Galactic polarized and unpolarized emissions in the microwave range : 
free-free, synchrotron, thermal dust and anomalous emission.  
\end{abstract}



\section{Introduction}

One of the greatest challenges in the observation of the Cosmic Microwave Background (CMB) 
in the 20-200~GHz range is its separation from the foreground emission. 
This task is facilitated by the fact that the intensity of the emission from 
the Galactic interstellar medium reaches a minimum in this range. 
On the other hand, even if the Galactic emission is weak, it is still stronger than
the CMB over a significant fraction of the sky.
In addition the identification of the cosmological signal is complicated by the fact 
that several interstellar emissions are superimposed in this frequency range: free-free, synchrotron, 
thermal dust. There are also strong evidences of an excess of emission in this range, discovered
by \cite{kogut1996,kogut1996a} in the COBE-DMR data. This so-called ``anomalous emission''
is remarkably well correlated with thermal dust emission at 100~$\mu$m
with a spectrum which can be confused with that of free-free or synchrotron emissions.

With a detailed characterization of the CMB unpolarized emission over the whole sky by COBE, WMAP
and soon Planck, most of the scientific focus is moving towards the observation of the
CMB polarized emission which will bring definite informations on the early Universe.
The study of the CMB polarized signal is instrumentally challenging but most importantly
the subtraction of Galactic foreground emission is tremendously more difficult than in unpolarized emission.
Even if the Galactic polarized emission is simpler (we expect only synchrotron and thermal dust to contribute
significantly to the Galactic polarized emission) the ratio of the CMB over Galactic polarization
is almost two order of magnitude smaller than in unpolarized light. 
It appears clearly that this task requires a deep understanding of the diffuse Galactic emission and
a coherent modeling of the unpolarized and polarized emissions.

\begin{figure}
\begin{center}
\includegraphics[width=\linewidth, draft=false, angle=0]{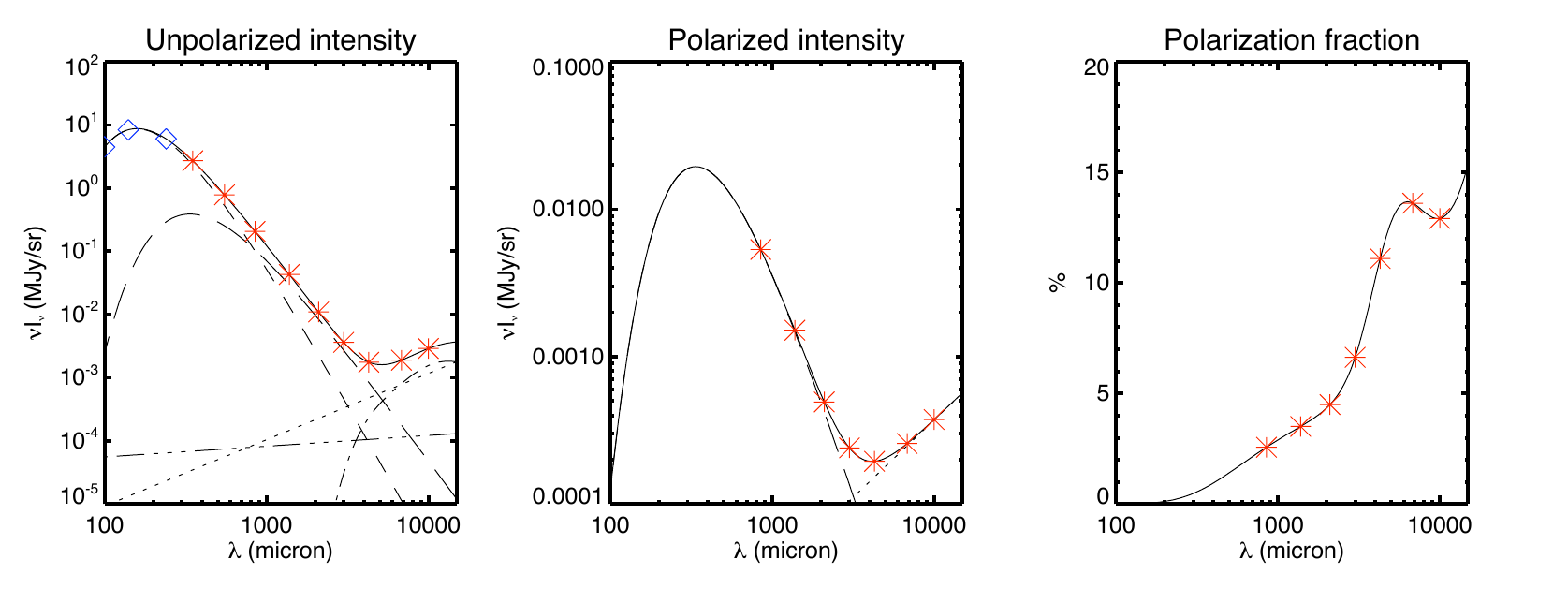}
\caption{\label{fig:foregroundSpectrum} Typical spectrum of diffuse emission at high Galactic latitude 
showing thermal dust (dashed), synchrotron (dotted), free-free (dash dot dot dot), spinning dust (dash dot) 
and total (solid).
Thermal dust is the sum of two components (\cite{finkbeiner1999} model 7): an unpolarized warm component
(short dashes : $T=16.4$~K, $\beta=2.60$ and $F_{pol}=0$\%) and a polarized colder one
(long dashes : $T=9.60$~K, $\beta=1.50$ and $F_{pol}=5$\%). The dust spectra
are only illustrative as the temperature of the two components in \cite{finkbeiner1999} are
significantly different than what is expected for interstellar dust \cite[]{draine2009}.
The synchrotron polarization fraction on this typical high latitude line of sight is 32\%. 
Free-free and spinning dust are assumed to be unpolarized.
Diamonds and stars indicate the frequencies of the DIRBE 
and Planck bands respectively.
The right panel shows the observed polarization fraction (i.e. polarized 
over unpolarized intensity). 
}
\end{center}
\end{figure}

Fig.~\ref{fig:foregroundSpectrum} shows a spectrum of the Galactic diffuse emission of a typical
line of sight at high Galactic latitude. The left panel 
illustrates the complexity of the superposition of
Galactic emissions in the CMB frequency range. In the infrared the emission
is dominated by the thermal emission from big dust grains.
Following \cite{finkbeiner1999} it is modeled here as the sum of two dust components
at different temperature (see \S~\ref{sec:thermaldust}). 
At longer wavelengths the power law spectra of the synchrotron 
and free-free emissions become dominant. Finally, the bell-shaped anomalous
emission appears at the lower end of the spectrum, with an intensity
that locally can dominate all the other components at some frequencies.
An attempt to separate all these contribution and produce all-sky maps 
for each of them has been presented by \cite{miville-deschenes2008}. 
The result of this decomposition for the five WMAP bands
is illustrated in Fig.~\ref{fig:mosaicforegrounds} where it is clearly seen that, 
in this range, there is no frequency where a single process dominates the emission. 

\begin{figure}
\vspace*{-5cm}
\hspace*{-1cm}
\includegraphics[draft=false, width=15cm, angle=0]{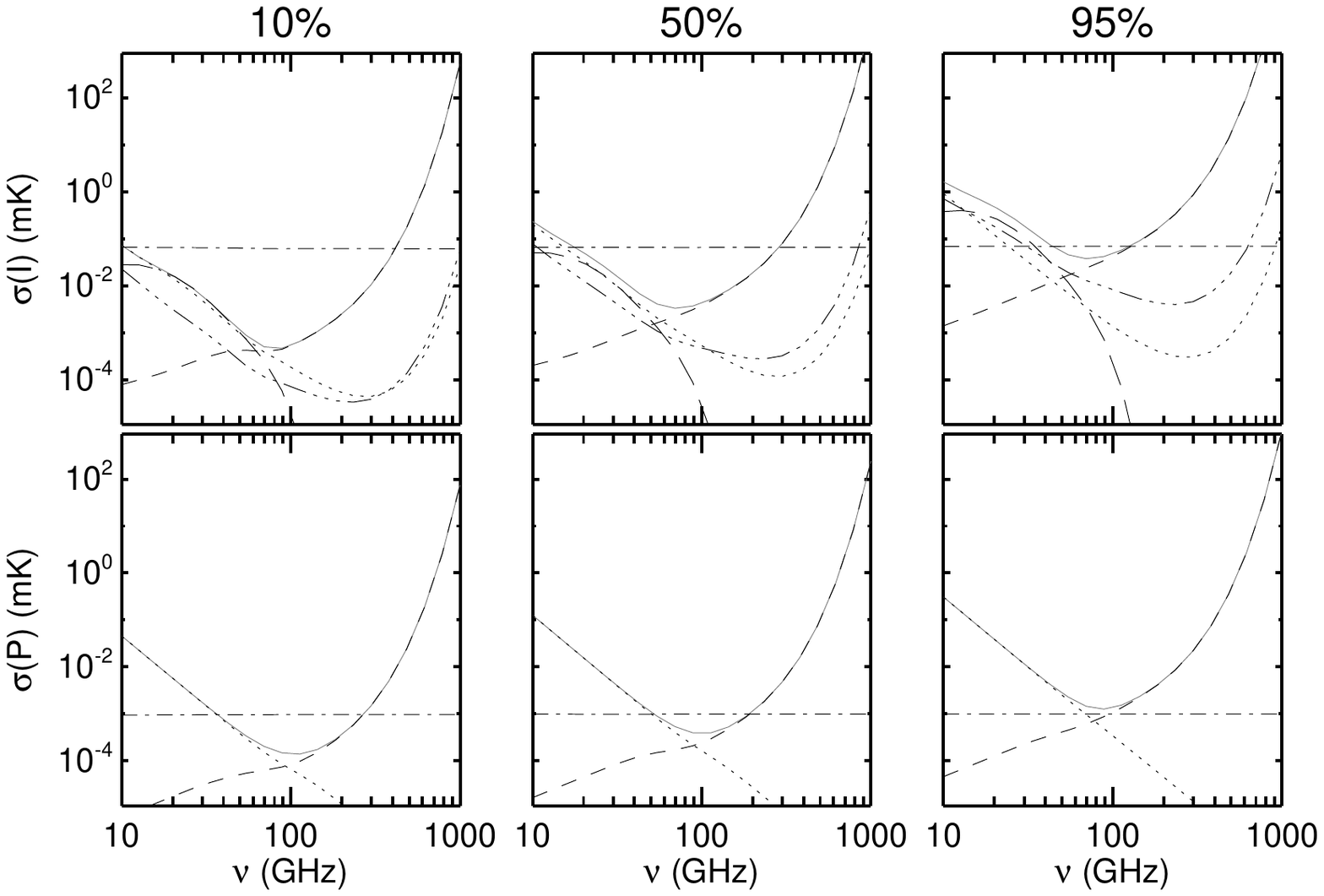}
\vspace*{-7cm}
\caption{\label{fig:spectre_ip} RMS fluctuations of each Galactic emission (synchrotron (dotted), 
free-free (dash dot dot dot), thermal dust (short dash), spinning dust (long dash), total (grey solid))
compared to the CMB (dash-dotted line) in units of mK (CMB), for sky coverage of 10, 50 and 95\%.
For each sky coverage we selected the $x$\% faintest pixels of the sky at each frequency - the mask is
therefore frequency dependent which explains the departure from pure power law of some components. 
Unpolarized emission is based on model~\#4 of \cite{miville-deschenes2008}. Polarized synchrotron is based
on the 23~GHz WMAP polarization data extrapolated in frequency with the same spectral index
as unpolarized emission. Polarized thermal dust is based on the unpolarized model of \cite{finkbeiner1999}
multiplied by a polarization fraction map based on the Galactic magnetic field model of \cite{miville-deschenes2008}
and normalized to the WMAP 94~GHz polarization data.
}
\end{figure}

Another way of appreciation the complexity of the emission in the CMB frequency range is by
looking at the rms fluctuation level of each emission as a function of frequency, for
different areas on the sky.
Fig.~\ref{fig:spectre_ip} shows this for polarized and unpolarized emission and for the
10, 50 and 95~\% faintest regions of the sky. It highlights the fact that, 
in regions of the sky of faintest Galactic emission, 
the CMB is clearly dominating the brightness fluctuations at frequencies around 80~GHz, but
the level of Galactic emission fluctuations increases greatly with the sky surface.
Fig.~\ref{fig:spectre_ip} also compares the level of fluctuation in polarization ($P=\sqrt{Q^2+U^2}$)
where the contrast of the CMB over the Galactic fluctuation level is much less favorable than in unpolarized light.
In this case the best frequency to observe the CMB is closer to 100~GHz.

The Stokes parameters Q and U can be express in terms of a scalar 
(curl-free) component of even parity (E modes) and a pseudo-scalar (curl) component of odd parity (B modes).
One of the greatest challenge of modern cosmology is to detect the B modes of the CMB polarization which
would provide key information on the existence of gravitational waves and on the inflation 
period of the Universe. These B modes are at least ten times
fainter that the E modes which dominates the polarization signal shown in Fig.~\ref{fig:spectre_ip}.
The detection of a non zero primordial B type polarization is one of the main challenges of 
actual and up-coming cosmological experiments (e.g. the Planck mission).
It will require a detailed description of the polarized Galactic emission, especially of polarized thermal dust.

This contribution is a review of the known constraints on the Galactic emission (polarized and unpolarized)
in the microwave range. The next sections are devoted to descriptions of the four main Galactic diffuse emissions :  
free-free (\S~\ref{sec:freefree}), thermal dust (\S~\ref{sec:thermaldust}), synchrotron (\S~\ref{sec:synchrotron})
and anomalous emission (\S~\ref{sec:anomalous}). 

\section{Free-Free}

\label{sec:freefree}

Free-free emission comes from the electron-ion scattering in the warm ionized medium at an average
temperature of $T_e\sim 8000$~K at high Galactic latitudes.
Its emission is a rather difficult component of the interstellar emission
to identify as it does not dominate at any frequency except in bright HII regions in the Galactic plane. 
H$\alpha$ corrected for dust extinction is often used as a proxy for free-free as they both depend linearly on the 
emission measure, EM=$\int n_e^2 \, dl$ (see \cite{dickinson2003} and \cite{finkbeiner2003}).
On the other hand, \cite{finkbeiner2004} argued that an error of a factor of 2 
is likely in some parts of the sky solely due to an imperfect extinction correction. 
In addition the estimate of free-free emission from H$\alpha$ requires an estimate of the electronic temperature which is 
usually assumed constant over the sky. This is a significant source of uncertainty as 
electronic temperature variations are expected, with lower values in the Galactic plane due to a higher metallicity
and therefore more efficient cooling.

\begin{figure}
\begin{center}
\includegraphics[height=11cm, draft=false, angle=270]{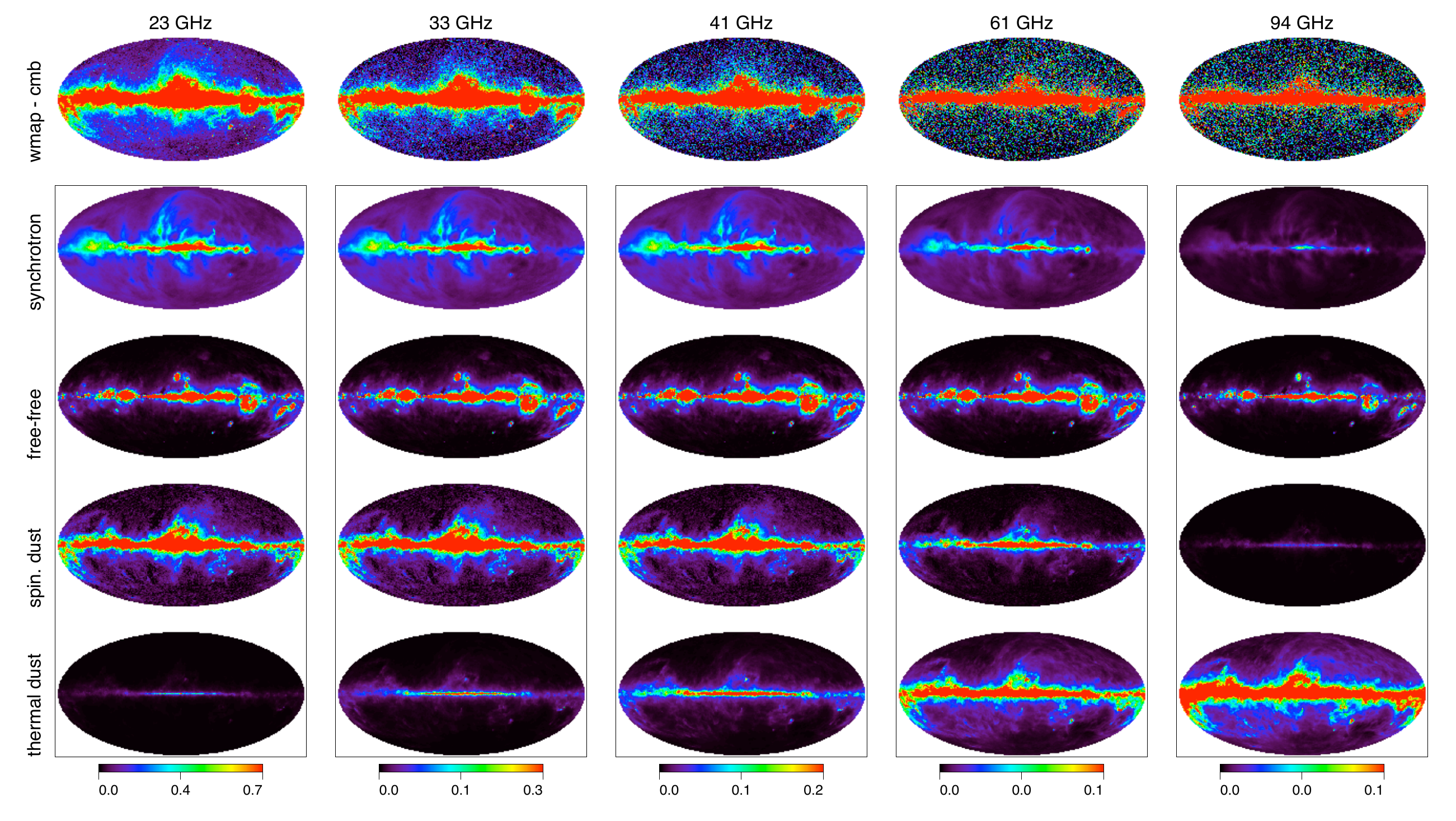}
\caption{\label{fig:mosaicforegrounds} Model of the Galactic emission (intensity only)
at WMAP frequencies. 
At each frequency is shown the WMAP (3 years) data, from which the CMB ILC map was removed,
and a decomposition of the Galactic emission in four components (synchrotron, free-free, spinning dust and thermal dust)
according to model~\#4 of \cite{miville-deschenes2008}. The dynamic range is the same for all images at 
a given frequency. All maps were smoothed to 1 degree. Units are mK (CMB).}
\end{center}
\end{figure}

It is important to point out that the free-free spectral index is a slowly varying function of frequency 
and electron temperature \cite[]{bennett1992,dickinson2003,bennett2003a},
At 23~GHz, the free-free spectral index varies only from 2.13 to 2.16 for $4000 < T_e < 10000$~K, independent of extinction.
Based on this physical property \cite{bennett2003a,hinshaw2007} estimated the
free-free emission in the WMAP data using a Maximum Entropy Method (MEM) considering a constant spectral index of 2.14
over the whole sky, providing an alternative to H$\alpha$.
\cite{miville-deschenes2008} built a new free-free template by combining these two estimates of the free-free emission; they used 
the sensitivity and resolution of the H$\alpha$ observations in regions where extinction by dust 
is limited \cite[]{dickinson2003} and the result of the WMAP MEM decomposition elsewhere \cite[]{bennett2003a}.
This map of free-free emission is shown in Fig.~\ref{fig:mosaicforegrounds}.

Free-free is essentially unpolarized, except at the edges of HII regions where Thompson scattering could produce
some polarization at small scales. \cite{dickinson2007} put an upper limit of only 0.6 percent for the polarized 
free-free emission of bright HII regions at 31 GHz. Therefore free-free is not expected to be a significant CMB foreground in
polarization.

\section{Thermal dust}

\label{sec:thermaldust}

\subsection{Intensity}

Interstellar dust grains have been known since the 1950s to produce extinction of starlight in the UV to near-infrared.
From the extinction curves it has been estimated that dust grains are made of silicate and carbonaceous material \cite[]{draine1984}
and that they have a size distribution that follows a power
law with sizes ranging from big molecules to about 1~$\mu$m \cite[]{mathis1977}. 
In the 80s the IRAS satellite observed the emission from these grains
over the whole sky at 12, 25, 60 and 100~$\mu$m with an angular resolution of 5 arcmin. 
These observations revealed the ubiquity
of interstellar dust and interstellar matter in general, even in the faintest part of the high Galactic latitude sky.

The near to mid-infrared part of the spectrum is produced by the emission of the smallest dust grains 
(radius $\leq 0.01 \mu$m)
heated stochastically by starlight.
As grains get bigger they absorb UV photons often enough to stop fluctuating in temperature and 
to be in thermal equilibrium with the ambient radiation field.
Therefore, in the far-infrared to millimeter range, the emission from big grains 
can be modeled as a modified black-body: $I(\nu) = \nu^\beta B_\nu(T_d)$ where $B_\nu(T_d)$ 
is the Planck function at frequency $\nu$, $T_d$ the equilibrium temperature and $\beta$ the emissivity spectral index.
The grain equilibrium temperature varies spatially as 
a function of the local radiation field and of the grains optical properties which depend on their 
chemical composition, size
and structure. For instance the growth of grains or the appearance
of icy mantles have an impact on their equilibrium temperature and on their emissivity parameter $\beta$ \cite[]{boudet2005,meny2007}.
The observed equilibrium temperature of big grains ranges from $< 10$~K in cold cores to several hundreds 
in photo-dissociation regions, with a typical value of 17.5~K ($\beta=2$) in the diffuse ISM \cite[]{lagache1998,desert2008}. 
The spatial and spectral variations of dust emission and extinction observed over the last 50 years revealed significant variations
in the dust properties (grain size distribution, structure and composition) which implies rather short evolution timescales 
through coagulation or fragmentation processes.

In the sub-millimeter and millimeter range the big grain emission has been measured by very few experiments. 
The FIRAS experiment on-board the COBE satellite made an all-sky map of this emission at a $7^\circ$ resolution. 
Molecular clouds have also been observed from the ground (e.g. JCMT at 450 and 850~$\mu$m - see \cite{johnstone2006}) 
or with balloon borne experiments like PRONAOS \cite[]{ristorcelli1998} or BLAST
\cite[]{chapin2008}. These observations revealed an excess of emission in the submm compare 
to the diffuse ISM spectrum ($T_d=17.5$~K, $\beta=2$). 
\cite{finkbeiner1999} showed that a better fit to the FIRAS data can be achieved by considering
two dust populations with different equilibrium temperatures and spectral indices.
Later, in the spirit of the work of \cite{draine1984} on the extinction curve, \cite{li2001}
also proposed a ``2 components'' model which, as opposed to the model of \cite{finkbeiner1999},
is also in accordance with known optical properties of silicate and carbonaceous dust.
It has also been proposed that the submm-mm dust spectrum could be explained by variations of the spectral index $\beta$
as a function of physical conditions. Laboratory measurements of absorption spectra
for amorphous silicates have revealed a change in $\beta$ in the submm range, as a function of $T_d$ \cite[]{boudet2005}.
These results are in accordance with the anti-correlation between $\beta$ and $T_d$ observed with the PRONAOS \cite[]{dupac2003}
and Archeops \cite[]{desert2008} experiments.
Significant progress on the exact nature of the big dust grains are expected soon with the Planck and Herschel missions.

\subsection{Estimate of thermal dust intensity in the CMB range}

Meanwhile, despite the fact that it is not coherent with extinction, the \cite{finkbeiner1999} provides a way
to estimate thermal dust emission in the CMB frequency range. These authors used the 100~$\mu$m/240~$\mu$m ratio (DIRBE data)
to estimate locally the average temperature of each dust component, assuming that the ratio of power absorbed (and radiated)
by each component is the same, and that the ratio of their abundance is the same everywhere. Assuming a
spectral index for each component, they extrapolate the 100~$\mu$m IRAS map to any frequency in the submm-mm range, providing
a template for thermal dust emission at 5~arcmin resolution.
Despite these rather strong hypothesis, this model reproduces rather well the WMAP 94~GHz data on scales larger than a few degrees.
On the other hand comparison at smaller scales with the 353~GHz data from the balloon-borne Archeops experiments \cite[]{benoit2004} 
seems to reveal differences up to 60\% (J.-P. Bernard, private communication).
A typical line of sight at high Galactic latitude 
for the model 7 of \cite{finkbeiner1999} is presented in Fig.~\ref{fig:foregroundSpectrum} (left panel).

\begin{figure}
\includegraphics[width=\linewidth, draft=false, angle=0]{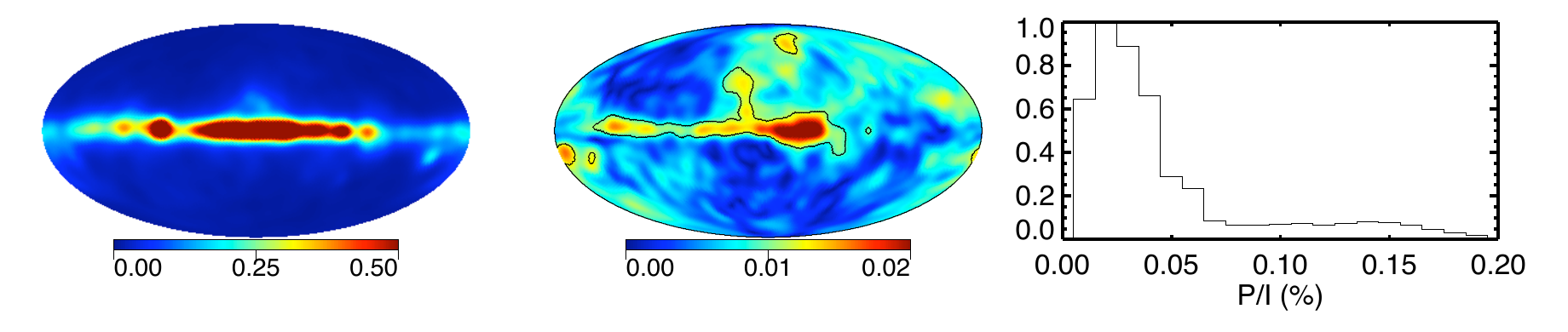}
\caption{\label{fig:polar94} 94 GHz unpolarized (left) and polarized intensity (middle) from the
WMAP 5 years data, smoothed at 10 degrees, in units of mK (CMB). 
Except in the Galactic plane, most structures seen in the polarization map
are due to noise in the Q and U Stokes parameters. The right panel shows an histogram of the
polarization fraction for regions where the signal-to-noise ration of the polarization data is significant (inside
the black contours in the middle panel). The polarization fraction is less than 6\% in the Galactic plane but it seems
to rise at higher Galactic latitudes.
}
\end{figure}

\subsection{Polarization}

Sub-millimeter polarized dust emission has been the subject of a number of review papers 
\cite[]{hildebrand2004,martin2007,vaillancourt2007} and of a detailed modeling study by \cite{draine2009}.
Extinction of starlight from dust in the UV to near-infrared is known to be polarized. This property reveals that dust grains
are aligned on the magnetic field lines, even if the exact mechanism by which they align is still not well understood 
(radiative torque is a likely candidate - see the review by Lazarian, this volume).
Polarized extinction has been used to estimate basic grain properties like shape, size and chemical composition \cite[]{bastien2007} but also
the structure and strength of the magnetic field \cite[]{heiles1996a}.

Given that the grains are aligned on the magnetic field lines, their emission in the far-IR/sub-mm is also polarized. 
From first principles we expect that a dust population with given optical properties
should have a polarization fraction ($P/I$) constant with frequency in the FIR to mm range \cite[]{hildebrand1988a}. 
Observations have shown that this is not the case in molecular clouds \cite[]{vaillancourt2002}; in this case
the wavelength dependence of the polarization fraction can be explained by the presence of more 
than one dust populations with different equilibrium temperatures. 
Considering the complexity of molecular clouds (in terms of magnetic field structure, variations of dust properties
and extinction of heating radiation field) it is far from clear if the same variation with wavelength
will be observed in the diffuse ISM. Up to now, the only observations of the polarized dust emission from the diffuse
emission in the submm/mm range were obtained at 353~GHz by Archeops \cite[]{ponthieu2005} and at 94~GHz by WMAP \cite[]{kogut2007}.
These studies show a clear detection at a few percent level (less that 6\% in the Galactic plane - see Fig.~\ref{fig:polar94})
with a spatial structure in the Galactic plane in accordance
with the starlight extinction polarization measurements. At 94~GHz the polarization fraction seems to increase slightly outside
the Galactic plane in accordance with a lower depolarisation effect due to the crossing of less magnetic field fluctuations 
on the line of sight than in the Galactic plane.
Unfortunately the sensitivity of these data is still not sufficient 
to allow a detailed analysis of the variation with wavelength of the polarization fraction it the diffuse ISM. 

The observation of such a variation would provide important constraints on the properties of dust
grains. Extinction measurements clearly showed that silicates type grains are aligned (O-Si band is polarized)
unlike carbonaceous type ones (carbon features are unpolarized). If big dust grains are composed of two distinct populations,
their different polarization fraction and equilibrium temperature will produce a clear spectral signature of the polarization
fraction in the submm, even in cirrus clouds at high Galactic latitudes (see Fig.~\ref{fig:foregroundSpectrum} - right panel - and
\cite{draine2009} for a modeling of this effect). 
This is a topic where the submm polarization observations from the Planck mission but also 
the balloon-borne experiments PILOT \cite[]{bernard2007} and Blast-pol \cite[]{marsden2008} will certainly 
make a break through in our understanding of interstellar dust.

\section{Synchrotron}

\label{sec:synchrotron}

The Galactic synchrotron radiation is produced by relativistic electrons spiraling around the Galactic magnetic field lines.
At a given frequency the synchrotron emission depends on the cosmic-ray electrons in a given energy range
and on the strength of the component the magnetic field perpendicular to the line of sight. The synchrotron
radiation is intrinsically strongly polarized.

\subsection{Intensity}

For a cosmic ray distribution following a power law, $N(E) \propto E^{-s}$, 
the synchrotron unpolarized intensity at frequency $\nu$ can be written as:
\begin{equation}
\label{eq:I_sync}
S(\nu) =  \int_z \epsilon_s(\nu) n_e B_\perp^{(1+s)/2} \, dz
\end{equation}
where the integral is over the line of sight $z$ and $B_\perp = \sqrt{B_x^2 + B_y^2}$
with the plane $x-y$ corresponding to the plane of the sky.
The emissivity term $\epsilon_s(\nu)$ is given by a power law:
\begin{equation}
\epsilon_s(\nu) = \epsilon_0 \nu^{\beta_s}
\end{equation}
where $\beta_s =-(s+3)/2$.
For a typical value of cosmic ray spectrum $s=3$ and $\beta_s=-3$.

The cosmic ray volume density $n_e$, the energy spectrum exponent $s$ and the $B$ field are
all varying locally in the Galaxy. 
Obviously, at a given frequency $\nu$ the synchrotron intensity varies on the sky, 
but the synchrotron spectral index $\beta_s$ also shows spatial variations which depends on local
variations of the cosmic rays energy spectrum.

It is worth noting that even if locally the energy spectrum is a power law, 
the fact that $s$ can vary along $z$ will induce departure from
a pure power law for the observed synchrotron spectral index.
In addition cosmic rays at higher energies (several GeVs)
lose their energy faster than low energy ones which produce a steepening of $N(E)$ at high energies
as we move away from cosmic-ray production regions.
This results in a steepening of the synchrotron spectrum at frequencies higher than $\sim$70~GHz,
which is expected to be stronger away from star forming regions where cosmic-rays are produced.
This effect is often modeled as a second order term in the synchrotron spectral index. 
It follows that the synchrotron intensity spectrum can be described as a power law
\begin{equation}
S(\nu) = S(\nu_0) \left(\frac{\nu}{\nu_o}\right)^{\beta_s+c \log(\nu/\nu_0)}
\end{equation}
where $c$ is a curvature term that could depends on sky position.

\begin{figure}
\includegraphics[width=\linewidth, draft=false, angle=0]{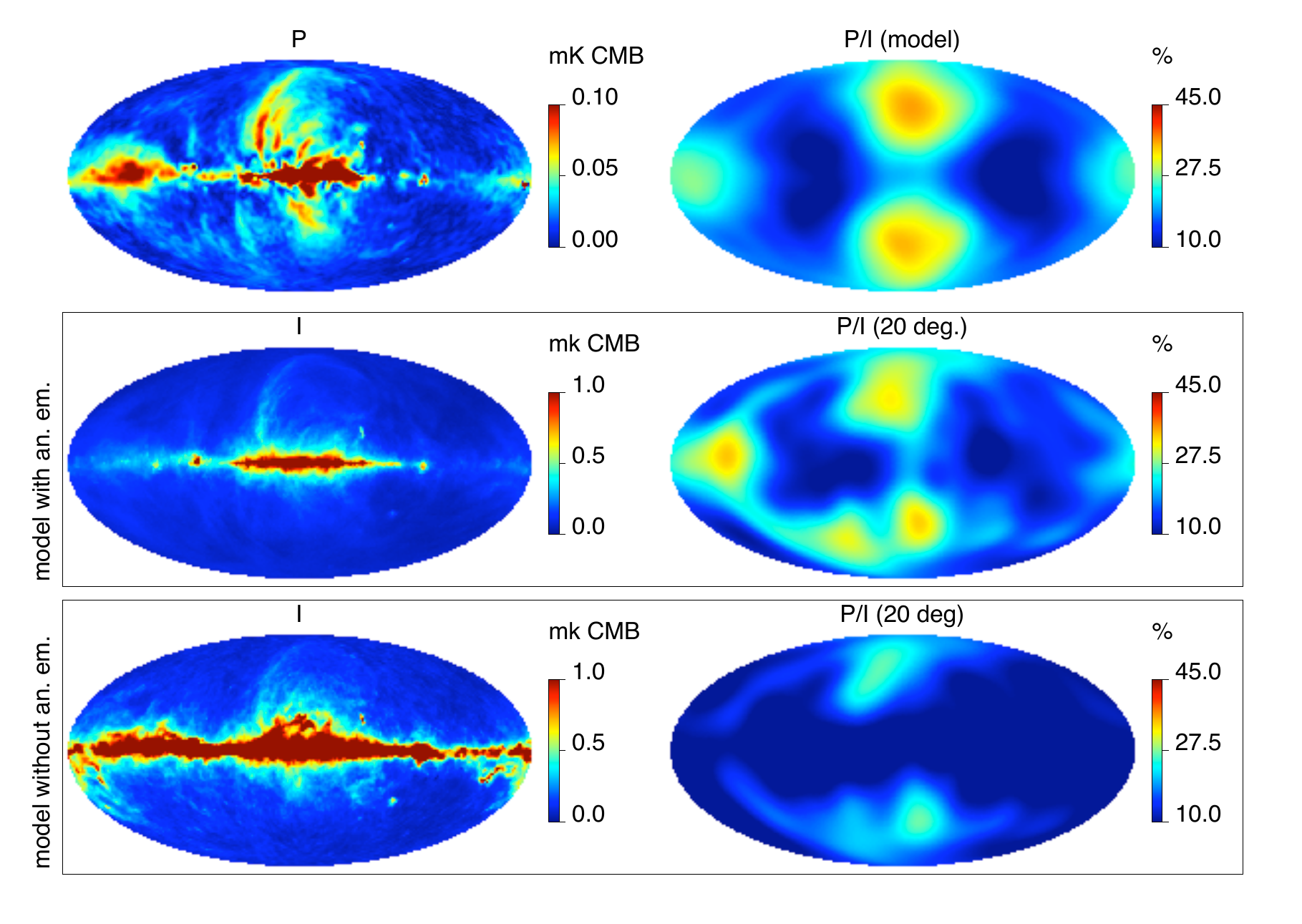}
\caption{\label{fig:modelsynchrotron} Synchrotron intensity and polarization at 23~GHz.
{\bf Top - left:} Polarized intensity as measured by WMAP at 23~GHz (smoothed at $5^\circ$).
{\bf Top - right: } Expected synchrotron polarization fraction at 23~GHz
based on a magnetic field decomposed in the sum of two component in energy equipartition :
a BSS large scale part and a turbulent part following a -5/3 power law at scales smaller than
100~pc.
{\bf Middle:} Model where the 23~GHz synchrotron intensity 
is estimated to be the Haslam 408~MHz intensity map extrapolated in frequency
with a constant spectral index of $\beta=-3$. In this model the 23~GHz emission is assumed
to be the sum of free-free, synchrotron, thermal and spinning dust. 
The right panel is the polarization fraction ($P/I$) for this model.
{\bf Bottom: } Model where the 23~GHz synchrotron intensity
is the WMAP 23~GHz data from which free-free and thermal dust were removed. In this model
the synchrotron intensity is higher (which explains the lower polarization fraction)
as it assumes that there is no contribution of the anomalous emission to 
the 23~GHz intensity.}
\end{figure}

\subsection{Polarization}

The synchrotron polarization fraction $f_s$ is related to the cosmic ray energy distribution slope $s$:
$f_s = (s+1)/(s+7/3)$.
For a typical value of cosmic ray spectrum $s=3$ the polarization fraction is 75\%.

This maximum polarization fraction is never observed on the sky because of depolarisation effects
due to beam depolarisation, to Faraday rotation, to the angle between the line of sight direction and 
the orientation of the large scale magnetic field lines, but mostly to the fluctuating direction 
of the random part of the magnetic field along the line of sight.

Similarly to the synchrotron intensity (see Eq.~\ref{eq:I_sync}), and neglecting Faraday rotation which is not important
in the microwave range,
the Stokes parameters $Q$ and $U$ of polarized synchrotron emission
integrated along the line of sight are 
\begin{equation}
\label{eq:Q_sync}
Q(\nu) = \int_z f_s \epsilon_s(\nu) n_e B_\perp^{(1+s)/2} \cos 2\phi \sin \alpha \, dz
\end{equation}
and
\begin{equation}
\label{eq:U_sync}
U(\nu) = \int_z f_s \epsilon_s(\nu) n_e B_\perp^{(1+s)/2} \sin 2\phi \sin \alpha \, dz
\end{equation}
where
$\cos 2 \phi = \frac{B_x^2-B_y^2}{B_\perp^2}$,
$\sin 2 \phi = \frac{-2B_xB_y}{B_\perp^2}$,
and
$\sin \alpha = \sqrt{1-B_z^2/B^2}$.

Like for the synchrotron intensity, the Stokes $Q$ and $U$ spectra can be approximated by power laws. On the other hand, 
because of the different weighting of the magnetic field in Equations~\ref{eq:Q_sync} and \ref{eq:U_sync} compared to
Eq.~\ref{eq:I_sync}, the spectral index of $Q$ and $U$ should have a slightly different structure on the sky.
In practice the determination of the spatial variation of the polarized synchrotron spectral index at CMB frequencies 
is challenging because of the limited signal-to-noise ratio of the WMAP data.

\subsection{Estimate of synchrotron intensity in the CMB range}

The most reliable estimate of synchrotron emission to date is the 408~MHz all-sky map of \cite{haslam1982}
(see middle-left panel of Fig.~\ref{fig:modelsynchrotron}).
The extrapolation of the 408~MHz emission at CMB frequencies relies on an estimate of $\beta_s$ and
of its variations on the sky (and eventually of the curvature term $c$).
All-sky estimates of the synchrotron spectral index between 408 MHz and 2326 MHz
have been attempted by \cite{giardino2002} and \cite{platania2003}. 
They obtained compatible values of $\beta_s = -2.7 \pm 0.1$ but slightly different
spatial variations. Unfortunately, due to cosmic-ray electrons aging effects \cite[]{banday1991}, 
these values of the synchrotron spectral index can not be used blindly to estimate synchrotron between 20 and 100~GHz.
The determination of the spatial variation of the 408~MHz-23~GHz synchrotron spectral index is made difficult by the 
presence of the anomalous emission. 
By looking at specific regions known to be dominated by synchrotron, 
\cite{davies2006} showed that it steepens to values of $\beta_s=-3.01\pm0.04$.
\cite{miville-deschenes2008} proposed a method to estimate $\beta_s$ on the whole sky
based on the use of the 23~GHz polarization data of WMAP. 

\begin{figure}
\begin{center}
\includegraphics[draft=false, angle=0]{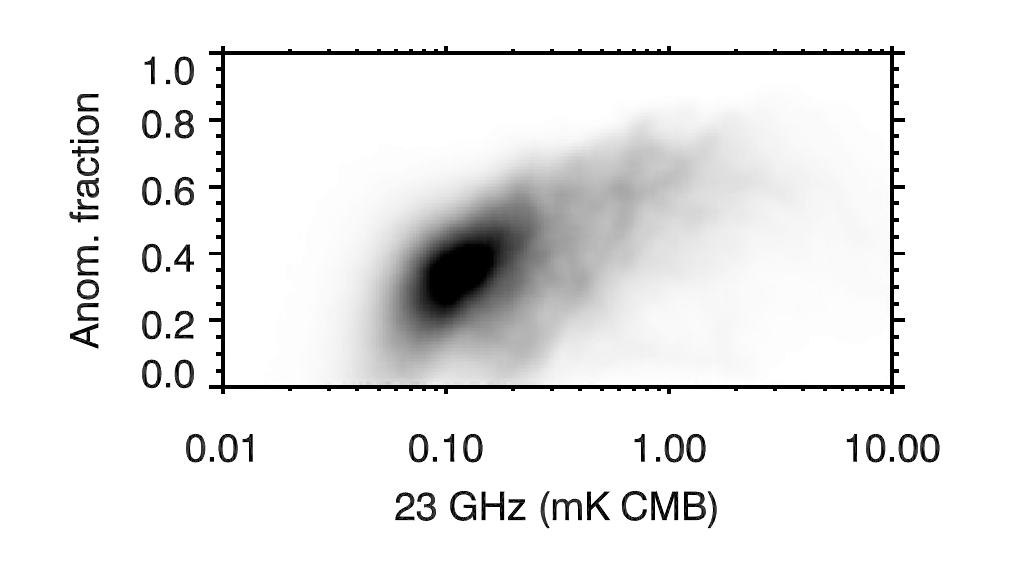}
\end{center}
\caption{\label{fig:anomalousFraction} Fraction of the total 23 GHz emission 
attributed to the anomalous emission as a function 
of total 23~GHz emission, according to \cite{miville-deschenes2008}.}
\end{figure}

\section{The anomalous emission}

\label{sec:anomalous}

\subsection{Synchrotron or spinning dust ?}

The analysis of the COBE-DMR data by \cite{kogut1996} revealed an excess of Galactic emission in the 30-90~GHz range, 
the so-called anomalous emission, strongly correlated spatially with thermal dust emission
but with a spectrum similar to synchrotron or free-free.
\cite{leitch1997} showed that the correlation of this excess with 100~$\mu$m extends to sub-degree scales
in the North Celestial Loop (NCL) region. The lack of correlation of the anomalous emission with H$\alpha$ measurements, 
also highlighted by \cite{de_oliveira-costa2002} using Tenerife and COBE-DMR data, 
led \cite{leitch1997} to conclude that free-free emission can only account for the data if it comes from gas with $T_e>10^6$~K.
\cite{draine1998} ruled out this interpretation based on energetic arguments.

An alternative explanation would be that the anomalous emission is actually synchrotron \cite[]{bennett2003,hinshaw2007}. 
As described by \cite{kogut2009} the high level of correlation of the 23 GHz data with thermal dust does not
provide a proof that the observed emission excess is not synchrotron, but this explanation 
needs to address the fact that the correlation of synchrotron with thermal dust would be stronger at 23~GHz than at 408~MHz.

In fact synchrotron emission observed at a given frequency $\nu$ is mostly produced by cosmic-ray electrons of a given energy $E$ following: 
\begin{equation}
\left(\frac{\nu}{\mbox{GHz}}\right) = 1.3 \times 10^{-2} \left(\frac{B}{\mbox{$\mu$G}}\right) \left( \frac{E_\nu}{\mbox{GeV}} \right)^2.
\end{equation}
Furthermore higher energy cosmic-ray electrons lose their energy faster than lower energy ones. Therefore the emission in the GHz range
produced by high energy CRs should be stronger closer to the regions where they are produced, i.e. star forming regions which
also have strong thermal dust emission. At MHz frequencies, the synchrotron emission comes from lower energy CRs which are more spread 
out in the Galaxy, explaining why the 408~MHz map is not strongly correlated with the thermal dust emission. 
As attractive as it might be this explanation suffers from several difficulties. 

First the correlation between the 23~GHz and the 100~$\mu$m maps extends to the resolution of WMAP ($\sim 1^\circ$ at 23~GHz)
and down to low column density cirrus clouds
\cite[]{davies2006} where there is no star forming activities and no local production of cosmic rays.

A second constraint comes from the low polarization fraction ($P/I$) observed at 23~GHz. 
Free-free is known to produce no polarization\footnote{expect at the edges of HII regions}. 
Once CMB and free-free are removed from the 23~GHz unpolarized emission $I$, 
the polarization fraction measured at 23~GHz is less than 10\% in the Galactic plane and 
it reaches 15\% on average at high Galactic latitudes \cite[]{kogut2007}.
As the intrinsic synchrotron polarization fraction is 75\% and that Faraday rotation is negligeable at 23~GHz,
this low polarization fraction requires either 1) the presence of an additional unpolarized emission or 
2) a strong depolarisation mechanism.
The most significant depolarisation process here is the tangling of the magnetic field lines along the line of sight.
Polarized extinction measurements in the near-infrared, pulsar measures and radio synchrotron observations all point
at a quasi-equipartition between the regular and turbulent part of $B$. 
Using a standard bi-symmetrical spiral model for the large scale magnetic field and considering a turbulent part of the 
field in energy equipartition with the regular part \cite{miville-deschenes2008} could estimate what should be the large scale 
polarization fraction map of synchrotron emission (Fig:\ref{fig:modelsynchrotron} top-right panel).
They conclude that in the case of equipartition, the modeled polarization fraction can be reconciled with the
observations only if an additional unpolarized component is present (Fig:\ref{fig:modelsynchrotron} middle panel). 
If not, as it was also shown by \cite{sun2008}, the polarization fraction observed at 23~GHz would require
a turbulent part of the magnetic field 1.5 times the regular part, which seems highly unlikely given the other constraints.
Once free-free and synchrotron (estimated from polarization) are removed from the 23~GHz unpolarized emission,
the residual emission is very well correlated to dust extinction E(B-V). \cite{miville-deschenes2008} could also 
estimated that this emission has an average spectrum between 23 and 61~GHz compatible with spinning dust models \cite[]{draine1998a}.

A third strong argument against synchrotron as the carrier of this anomalous emission came recently from the analysis
of the absolutely calibrated ARCADE data at 3, 8 and 10~GHz.
\cite{kogut2009} showed that combined with the Haslam 408~MHz map the ARCADE data put a very strong constraint on the synchrotron spectral index 
near the CMB frequency range. The extrapolation of these data to 23~GHz clearly shows 
that synchrotron plus free-free can not account for all the 23~GHz Galactic emission
and that the residual emission is of the order of $40\pm10$\% of the 23~GHz emission in the Galactic plane.
This is in accordance with the analysis of \cite{miville-deschenes2008} who conclude 
that the fraction of the 23~GHz emission attributed to the anomalous emission 
is about 30\% at high latitudes and of 50\% on average in the Galactic plane (see Fig.~\ref{fig:anomalousFraction}).

\subsection{Polarization of the anomalous emission}

The question of the polarization of the anomalous emission is an important one in the context of the analysis of the
polarized CMB emission.
Polarization of the dust-correlated anomalous emission has been looked for 
around 10~GHz in specific clouds by \cite{battistelli2006} and \cite{mason2009}. These
studies find marginal polarization fraction for the anomalous emission. These results allow 
to exclude magnetic dipole models \cite[]{draine1999} and are in favor of
the spinning dust emission \cite[]{draine1998a} which is the result of rotational excitation
of the smallest dust grains (i.e. PAHs). The polarization extinction measurements in the UV clearly show
that these small grains are weakly aligned on the magnetic field,
unlike bigger grains. Therefore the spinning dust emission is not expected to be significantly polarized.
Nevertheless, as the level of alignment of the PAHs on the magnetic field is not well constrained,
it is not clear if polarized spinning dust from the diffuse ISM is present or not in the 10-100~GHz range.
Its detection is challenging because of its low polarization level, the strong
polarization fraction of the synchrotron emission and the limited signal-to-noise ratio of the 
available data. 

\section{Conclusion}

The Galactic emission in the microwave range is rich and complex. Even with only continuum emission
(free-free, synchrotron, thermal dust and anomalous emission),
this wavelength range provides important diagnosis and tracers 
for all phases of the interstellar medium, cosmic rays, dust grains and the Galactic magnetic field. 
Fortunately for research on the interstellar medium, the detailed understanding of the Galactic 
microwave emission is one of the main challenges shared by all CMB experiments as it is proven
to be the main obstacle for its study, especially in polarization.

It appears that the understanding of interstellar dust, through the thermal emission from big grains
and the rotational emission from smaller ones, will be a major aspect of the CMB data analysis.
Interstellar dust have been studied for a long time in the infrared and in UV-NIR extinction.
Nevertheless very little is known on the big grains microwave emission, especially in polarization,
but yet enough to expect significant spatial and spectral variations.
With its adapted frequency coverage the Planck mission will provide the first all-sky map
of dust polarized emission in the submm and, with it, a totally new view of the Galactic magnetic field.

At lower frequencies a strong debate about the existence of an anomalous emission
and its separation from synchrotron has been on-going since the COBE results.
It seems now clear that there is an extra component in the 20-50~GHz that can not be attributed
to any known process. This emission has a very low (if any) polarization and it is very well
correlated to interstellar dust tracers. Emission from small spinning dust grains is
a very likely candidate.

Significant progress have been made since the COBE era on the 
understanding of the Galactic emission in the microwave range
but the detailed exploration is only beginning. 
High hopes are set on on-going and future projects, especially the Planck mission. 

\acknowledgements 
I would like to thank the conference organizers for inviting me. 
I acknowledge funds from the Canadian Space Agency in support of this work.

\bibliographystyle{/Users/mamd/lib/tex/style/aa-package/bibtex/aa}
\bibliography{/Users/mamd/BIB/all.bib}


\end{document}